# Counterfeits on Dark Markets:
# A measurement between Jan-2014 and Sep-2015

Felix Soldner[1,2,*], Bennett Kleinberg[1,3], Shane D Johnson[1]

[1]Department of Security and Crime Science & Dawes Centre for Future Crime,
University College London, UK

[2]GESIS – Leibniz Institute for the Social Science, Köln, Germany

[3]Department of Methodology and Statistics, Tilburg University, The Netherlands

*Corresponding author: felix.soldner@ucl.ac.uk

Abstract

Counterfeits harm consumers, governments, and intellectual property holders. They accounted for 3.3% of worldwide trades in 2016, having an estimated value of $509 billion in the same year. Estimations in the literature are mostly based on border seizures, but in this paper we examined openly labeled counterfeits on darknet markets, which allowed us to gather and analyze information from a different perspective. Here, we analyzed data from 11 darknet markets for the period Jan-2014 and Sep-2015. The findings suggest that darknet markets harbor similar counterfeit product types to those found in seizures but that the share of watches is higher while the share of electronics, clothes, shoes, and Tobacco is lower on darknet markets. Also, darknet market counterfeits seem to have similar shipping origins as seized goods, with some exceptions, such as a relatively high share (5%) of dark market counterfeits originating from the US. Lastly, counterfeits on dark markets tend to have a relatively low price and sales volume. However, based on preliminary estimations, the equivalent products on the surface web appear to be advertised for a multiple of the prices found for darknet markets. We provide some suggestions on how information about darknet market counterfeits could be used by companies and authorities for preventative purposes, showing that insight gathering from the dark web is valuable and could be a cost-effective alternative (or compliment) to border seizures. Thus, monitoring darknet markets can help us understand the counterfeit landscape better.

Analysis code and supplementary data can be found at: https://osf.io/32au4/

## 1. Introduction

Counterfeits are illicit goods that violate intellectual property (IP) rights such as copyrights, trademarks, design rights, or patents, and they can exist physically or digitally (OECD/EUIPO, 2019; WTO, 1994). The purpose of a deceptive counterfeit is to make a monetary profit by deceiving a customer that the product is of a higher value than it is (e.g., by selling it as being genuine) (OECD/EUIPO, 2019). Deceptive counterfeits can be sold directly to consumers or mixed within supply chains of genuine products to reduce costs and increase profits (Hollis & Wilson, 2014). Counterfeits can cause a variety of problems, such as physical (e.g., through foods or pharmaceuticals) and monetary harms to the consumer, the IP holder (e.g., through damages to the brand value, loss of sales), or the government (e.g., through the loss of tax income) (EMCDDA-Europol, 2017; OECD/EUIPO, 2019). In turn, the sales of counterfeits can support organized crime groups financially and facilitate other illegal activities, such as money laundering (EMCDDA-Europol, 2017; UNICRI & ICC BASCAP, 2013; UNODC, 2014). The OECD/EUIPO (2019) estimated that counterfeits made up 3.3% of worldwide trades in 2016, worth USD 509 billion. Furthermore, the proportion of counterfeits seems to be elevated within developed regions, such as the European Union (EU).

However, estimating counterfeit goods' trade (value) is difficult and is mostly achieved through auditing goods seized at borders (OECD, 2018; OECD/EUIPO, 2019). Thus, current estimates often exclude domestically traded counterfeits or digital products, and since not all counterfeits will be seized at ports, estimates of what is traded may be incomplete. For example, the number of routinely checked containers at major ports in Genoa (Italy), Melbourne (Australia), Montreal (Canada), New York (USA), and Liverpool (UK) together only account for 2-5% of all traffic (Sergi, 2022). Since only a limited number of containers can be checked, the selection procedure can strongly impact possible finds.

A theoretical and empirical understanding of how counterfeiting occurs is currently not well developed, perhaps due to the complex involvement of various stakeholders, which results in difficulties for researchers to obtain reliable data (Sullivan et al., 2017). For example, many companies affected by counterfeiting operate across nations, affecting the ease with which authorities can monitor and combat counterfeits. Moreover, the definition of counterfeits varies across nations, further complicating how counterfeiting is measured. However, theories provide perspectives as to why counterfeiting occurs and how it might be addressed. The Rational Choice perspective considers the offender's choice to commit a crime (e.g., counterfeiting a product) and influencing factors of the offenders' decisions, such as the perceived risks and rewards (Clarke & Cornish, 1985). The perspective informs us that while changing the perceived risks and rewards of the offender, the likelihood of offending can be altered and reduced, for example, by increasing the perceived risks of detection or by increasing the general efforts needed to commit a crime. Within the context of counterfeits facilitating the traceability of genuine products within a supply chain (e.g., through watermarks) seems to be a possible approach to increasing the efforts to counterfeit (Gayialis et al., 2022). Another perspective, such as the Routine Activity Approach (RAA), discussed by Spink et al. (2013, 2014), states that crime is more likely when a suitable target (e.g., a product that can be counterfeited) and a motivated offender converge absent a capable guardian (Cohen & Felson, 1979). Capable guardians can include those involved in security at country borders or those involved in inspecting goods at other stages of the supply chain (Marucheck et al., 2011; Tang, 2006). For example, when manufactured products are transported, transport personnel and employees could also act as guardians (Hollis & Wilson, 2014). However, effective guardianship requires a clear understanding of the problem and processes to monitor it, such as reporting procedures. Absent this understanding, guardianship will be less effective.

With this in mind, risk assessments are often conducted to aid decisions made by authorities at borders based on intelligence from federal and local authorities and custom officer experiences (Sergi, 2022). However, these are likely to be imperfect. Furthermore, border checks can be random or may only be informed by the country of origin, or how the delivery is labeled, as in the case of parcel shipments (Männistö et al., 2021). Another source of possible bias present in check-selection procedures are large differences in the estimates of counterfeit product types produced by different agencies (IP Crime Group, 2015; OECD/EUIPO, 2019). For example, estimates strongly differ across agencies for footwear (20 percentage points) or for electronics (11%) and clothing (10%). For other products, estimations may be missing entirely, as in the case of Tobacco, which was estimated to make up 28% of all counterfeits by the IP Crime Group (2015) but was not identified as a counterfeited product by OECD/EUIPO (2019). Since different agencies use different data sources (e.g., border or inland seizures), some measurement differences are to be expected, but they also illustrate how inconsistently seizures reflect the true prevalence of counterfeits. Thus, additional data sources to estimate counterfeit affected products would be helpful to better understand the counterfeit landscape and aid efforts at prevention.

With the emergence of dark markets which do not have formal guardians in the same way that open web platforms do, new ways of trading illicit goods, including counterfeits, have appeared (Christin, 2013; van Wegberg et al., 2018), which may serve as an additional data source to measure counterfeit prevalence. Dark markets are online shopping platforms on the deep web – a highly anonymized part of the internet that is not indexed by traditional search engines – which operate like their surface web counterparts, eBay or Amazon. Vendors on dark markets offer a range of illegal products and services, mainly consisting of drugs, but also including hacking services, weapons, guides on how to defraud people, and counterfeits (Baravalle & Lee, 2018; Roberts & Hernandez-Castro, 2017; Soska & Christin, 2015; van Wegberg et al., 2018). During the COVID-19 pandemic, dark markets also started to offer a mix of genuine and fake protective gear (masks, gloves, etc.), medicines, and COVID-19 vaccines (Bracci et al., 2021a, 2021b; Broadhurst & Ball, 2020). Even with successful disruptions and the closing of markets by law enforcement, dark markets increasingly trade in such products and services (Décary-Hétu & Giommoni, 2017; ElBahrawy et al., 2020; EMCDDA-Europol, 2017). On dark markets, vendors openly sell counterfeits and forgeries, which provides an interesting opportunity to gain insight into the counterfeit market from a new angle. Since some dark markets also register the number of goods sold and buyers leave reviews, we can use such information to generate estimates of sales volumes and the monetary value of counterfeits over time. By comparing counterfeit listings and their sales on dark markets to border seizures, we can also see if they differ and provide a more comprehensive picture, which would be of value to law enforcement, companies that are affected by counterfeits of their products, and policymakers.

Therefore, to better understand the counterfeit economy on the dark web, we examined the prevalence and sales of counterfeits sold on 89 dark markets for the three-year period January 2014 – January 2017. Specifically, we quantified the price, volume, type, and origins of advertised counterfeits and estimated their sales volume and the value the same counterfeits would attract on the surface web. We then compare the results to measures and estimations from border seizures conducted by law enforcement over the same period. By highlighting differences, we can identify product groups for which counterfeiting appears to be a problem and would be overlooked based on an analysis of seizures alone.

### 1.1. Fraud and counterfeits on dark markets

Studies that have investigated the types of products listed and sold on the dark web mostly cover illegal drugs, which often account for 60-80% of all listings on a dark market (Baravalle & Lee, 2018; EMCDDA-Europol, 2017). However, some studies have examined less frequently listed products, such as art, wildlife, and plane tickets (Hutchings, 2018; Paul, 2018; Roberts & Hernandez-Castro, 2017). Others focus on fraud-related products or services, such as credit card information, online accounts (e.g., e-bay), social engineering guides and tutorials, or financial malware (e.g., ransomware) (Garg et al., 2015; Marin et al., 2016; Schafer et al., 2019; van Wegberg et al., 2018). Although some of these studies have considered the sale of forged documents (e.g., passports, licenses, diplomas), none have investigated or quantified the sales of counterfeits in a systematic way, such as differentiating between clothing, shoes, electronics, or jewelry; product types which can also be found on surface web markets (e.g., eBay, Amazon). Europol (2017) draws attention to IP crime on the dark web and estimates that solely counterfeit goods make up around 1.5-2.5% of all listings on such markets. The report lists some of the types of counterfeits sold on darknet markets (e.g., clothes, accessories, electronics, jewelry, pirated goods), and discusses the presence of wholesalers, which seem to account for the minority of transactions but most of the sales volume. In contrast, they report that most of the transactions are through the sales of individual items but seem to account for the minority of sales volume. According to this report, counterfeits seem to be sold for 1/3 of the price of the equivalent genuine product, and digital goods for around 1/6 of their original price (Europol, 2017). The report concludes that while the sale of IP goods is limited, there is potential for growth on darknet markets, and IP goods on dark markets should be monitored and investigated in more detail. However, the report does not explain how the mentioned statics were obtained nor which darknet markets were included in the analyses, making it difficult to assess the extent of counterfeits on the dark web. Furthermore, the lack of granularity prevents us from understanding which product types are offered, how frequently, how much they are sold, and where they originate. Lastly, the Europol (2017) report does not differentiate between counterfeits that could be sold on the surface web (e.g., shoes, clothes, electronics) and counterfeits that are limited to the dark web (e.g., fake banknotes or IDs), which is important if we want to inform authorities or companies on potentially affected product types that could be sold on the surface web.

### 1.2. Aims of this paper

With this paper, we aim to address the shortcomings of previous work by examining an extensive collection of dark market datasets to (**I**) understand the prevalence of counterfeit goods on the dark web and (**II**) determine the product types, occurrences, and origins of the identified counterfeits. Determining those details will help us (III) report counterfeit prices more accurately (by product types) and make sales volume estimations through product feedback, which can help us better understand the counterfeit economy on the dark web. Subsequently, we (**IV**) compare dark web counterfeit prices with prices of the same products on the surface web to understand possible profit margins for the various product types identified. We then (**V**) compare our results to observations made through border seizures, complaint statistics, and activities from authorities to contribute to the overall understanding of the counterfeit economy. Lastly, we (**VI**) discuss our results in relation to theoretical perspectives to provide future research avenues and possible implications for prevention or intervention approaches for authorities and companies facing counterfeits.

### 2. Data

The data used in this study originated from the "Darknet Market Archive"[1], a collection of 89 markets and associated forums (Branwen et al., 2015) for which data were initially collected between 2014-2015 and continuously supplemented thereafter. To facilitate the selection of relevant markets, we cross-referenced the available market data with a list of markets documented by EMCDDA-Europol (2017). Through this comparison, we identified 38 markets (see Appendix A), each of which operated for at least six months and was captured in the data archive. The reason for Including markets that operated for at least six months was to ensure that the markets were able to attract enough vendors and customers, allowing for a broader range of product offers and trades[2]. The market archive contained data on 30 of the 38 identified markets, but five of them contained data spanning less than six months, and data on eight markets did not include a sufficient self-organizing structure (e.g., categories), which would have allowed for the identification of counterfeit goods. For example, some market data contained products (e.g., shoes, handbags) without categorization or a detailed description, making it impossible to determine if they were counterfeits or originals that had been stolen or otherwise illegally obtained. Furthermore, six markets were either highly specialized (e.g., solely carding or Marihuana markets) or did not contain any counterfeits. Thus, we included the remaining 11 markets in our study (see Table 1).

---

[1] Data: https://www.gwern.net/DNM-archives

| Name | Data-start | Data-end |
|---|---|---|
| The Marketplace | 03.01.2014 | 09.11.2014 |
| Agora | 01.01.2014 | 07.07.2015 |
| Evolution | 21.01.2014 | 17.03.2015 |
| Cloud 9 | 11.02.2014 | 01.11.2014 |
| BlackBank Market | 06.02.2014 | 17.05.2015 |
| Andromeda | 12.04.2014 | 18.11.2014 |
| Middle Earth Market | 23.06.2014 | 05.07.2015 |
| Diabolus/Silk Road 3 | 17.10.2014 | 05.07.2015 |
| Abraxas | 16.12.2014 | 05.07.2015 |
| AlphaBay | 21.12.2014 | 28.01.2017 |
| Crypto Market | 19.02.2015 | 06.07.2015 |

Table 1. Markets and their data timeframe in this study.

### 2.1. Data filtering

Each market listed a range of products that were not counterfeits (e.g., drugs, services, weapons). Consequently, it was necessary to exclude such listings prior to analysis. To do this, we created a corpus of counterfeit products in two steps[3]. First, we included products that were clearly categorized as counterfeits based on the categories used on the markets, such as "Counterfeit[s]", "Replica[s]", "Counterfeit Items", and "Replica watches". Second, listings that were not included on this basis were filtered using an advanced keyword search. These keywords were for 29 other categories of items that, through the manual inspection of the data, were identified as including counterfeits (see Appendix B for the complete list of the categories). To facilitate the advanced keyword search, we merged the title and description of each listing in those 29 categories, lowercased, tokenized, and stemmed the text, and removed all punctuation. We then searched for 44 stemmed synonyms of the word "counterfeit" (e.g., "fake", "clone"; a complete list is provided in Appendix C) as well as six negated synonyms of "authentic", using bigrams (e.g., "genuine", "original"; a complete list is provided in Appendix D) in each merged listing text. Lastly, a list of keywords was used to exclude listings (Appendix E) that sold templates or tutorials on how to counterfeit. 124,379 listings were clearly marked as counterfeits, while 42,775 listings were identified through the keyword searches, resulting in a total of 158,228 counterfeit listings overall. Of these, 11,633 were completely unique listings for which at least the title, description, and vendor name differed. Text processing was conducted using the python package "nltk" (Bird et al., 2009).

---

[2] Manual inspections of the data revealed that markets with a shorter lifespan did not sell any counterfeits and often harbored very few vendors.
[3] The code used to process and analyze the data can be found at: https://osf.io/32au4/

## 2.2. Categorizing counterfeits

To determine the distribution of product types among those identified as counterfeits, we trained a machine-learning classifier on a subset of human-annotated data. The classifier was then used to predict the categories of the remaining unannotated products. To generate the annotated data, we randomly extracted 2200 unique listings from the counterfeit data set, which participants from the crowdsourcing platform *Prolific* subsequently annotated[4]. To ensure that we obtained accurate annotations, each listing was annotated by at least three participants. We recruited 220 participants, each annotating 30 listings based on the listing title. Participants provided written informed consent online by clicking all consent statement boxes affirming their consent before taking part in the study. The final category label for each listing was determined using the majority vote.[5] When annotating, participants were presented with an online interface and were required to select one of the following labels: "Watches", "Handbags", "Wallets", "Sunglasses", "Other accessories", "Clothing", "Footwear", "Articles of leather", "Fabrics (silk, rugs)", "Phones", "Electronics", "Jewelry", "Cosmetics", "Pharmaceuticals", "Metals", "Tobacco", "Forgeries (Money, Coupons, IDs, etc.)", "Services", "Other".[6] We calculated Krippendorff's alpha to determine how much annotators agreed on the labels they generated[7] (Feng, 2015). The value of α = 0.75 demonstrated good agreement (Hayes & Krippendorff, 2007; Krippendorff, 1970).

| Category | # | % |
|---|---|---|
| Watches | 902 | 40.76 |
| Forgeries | 168 | 7.59 |
| Clothing | 166 | 7.50 |
| Other | 148 | 6.69 |
| Footwear | 138 | 6.24 |
| Electronics | 121 | 5.47 |
| Handbags | 120 | 5.42 |
| Sunglasses | 108 | 4.88 |
| Jewelry | 81 | 3.66 |
| Other accessories | 69 | 3.12 |
| Services | 64 | 2.89 |
| Wallets | 34 | 1.54 |
| Metals | 34 | 1.54 |
| Pharmaceuticals | 22 | 0.99 |
| Cosmetics | 21 | 0.95 |
| Tobacco | 17 | 0.77 |

Table 2. Annotated categories within counterfeits

From the distribution of categorized products, it was apparent that the product types were not uniformly distributed, with watches representing the majority of all counterfeits annotated. Because some of the categories had low numbers, which would likely affect the classifier's performance, when training the classifier, we manually added eight listings to the "Tobacco" category and six listings to the "Cosmetics" category. Table 2 shows the resulting distribution (after manually adding listings) of the labeled categories for the randomly selected subset of counterfeits.

---

[4] www.prolific.co
[5] Due to the allocation procedure of participants from Prolific to our annotation task, some listings were only annotated twice while some were annotated more than three times. 192 annotations ties were manually resolved by the first author.
[6] The categories were determined based on reported counterfeits for seized goods by law enforcement (OECD/EUIPO, 2019).
[7] Krippendorffs alpha ranges between -1 (perfect disagreement) and 1 (perfect agreement) and can account for unequal numbers of annotators and annotations per item as well as missing annotations.

### 2.3. Automated labeling

Inspired by previous research (van Wegberg et al., 2018), we used the annotated listings to train a multiclass classifier to predict the labels of the remaining unlabeled counterfeits. Obtaining labels for all the listings has the advantage of allowing us to conduct our analyses for the whole dataset, including the price or individual texts of the listings, which would be more difficult through estimations from a sub-sample. We generated text features from the merged product title and description to train the classifier. However, we first lowercased the texts and removed all punctuation. We then tokenized the text, removed all English stop words, and stemmed the remaining words. Subsequently, we generated part of speech tags, unigrams, and bigrams, which were weighted with a tf-idf (term frequency-inverse document frequency) score. The python package "nltk" (Bird et al., 2009) was used for all text cleaning and feature generation steps. To increase the classifier's performance, we used a mix of under- and over-sampling methods to balance the number of product listings between the categories. First, the category "Watches" was under-sampled, reducing the number of listings in the sample. This was followed by oversampling of the remaining categories to increase the number of these listings in the sample, resulting in an equal representation between all categories, each consisting of 450 listings. To reduce the number of listings within each category, we randomly selected listings (without replacement) from the data until we reached 450 listings. To increase the number of listings within a category, we used "SMOTE" (Synthetic Minority Oversampling Technique), which synthesizes new unseen data points (Chawla et al., 2002). Such new data is generated by first randomly selecting a listing of that category and finding the $k$ (5) nearest neighbors of that listing within the feature space. Then, one of the neighbors is selected at random, and a new data point is created at a random point between the two listings in their feature space. Both under- and over-sampling methods were implemented in python using the package "imblearn" (Lemaître et al., 2017). Next, we utilized the "LinearSVC" classifier with an "l2" penalty (the default regularization parameter used to reduce complexity in the model and avoid overfitting) using a 10-fold cross-validation procedure. The under-, over-sampling, training, and testing steps were embedded within a pipeline so that the classifier was trained on the balanced listings (450 in each category) but tested on the unbalanced listings (as in table 2), ensuring a fair assessment. The test performances were evaluated using the average accuracy, and the weighted average of precision, recall, and F1 scores across all folds, as shown in Table 3. The python package "scikit-learn" (Pedregosa et al., 2011) was utilized for training, testing, and evaluating the classifier.

| Accuracy | Precision | Recall | F1 |
|---|---|---|---|
| 0.85 | 0.85 | 0.85 | 0.85 |

Table 3. The performance scores (weighted average) across 10-folds.

To better understand the classifier's performance for each category, we generated a normalized confusion matrix for all classes (Fig. 1). The matrix shows the cases of true (rows) and predicted (columns) categories of the listings. Thus, the values in the matrix show the proportion of items for which the true class was predicted. The diagonal cells (left-top to right-bottom) indicate the correct proportion for each category.

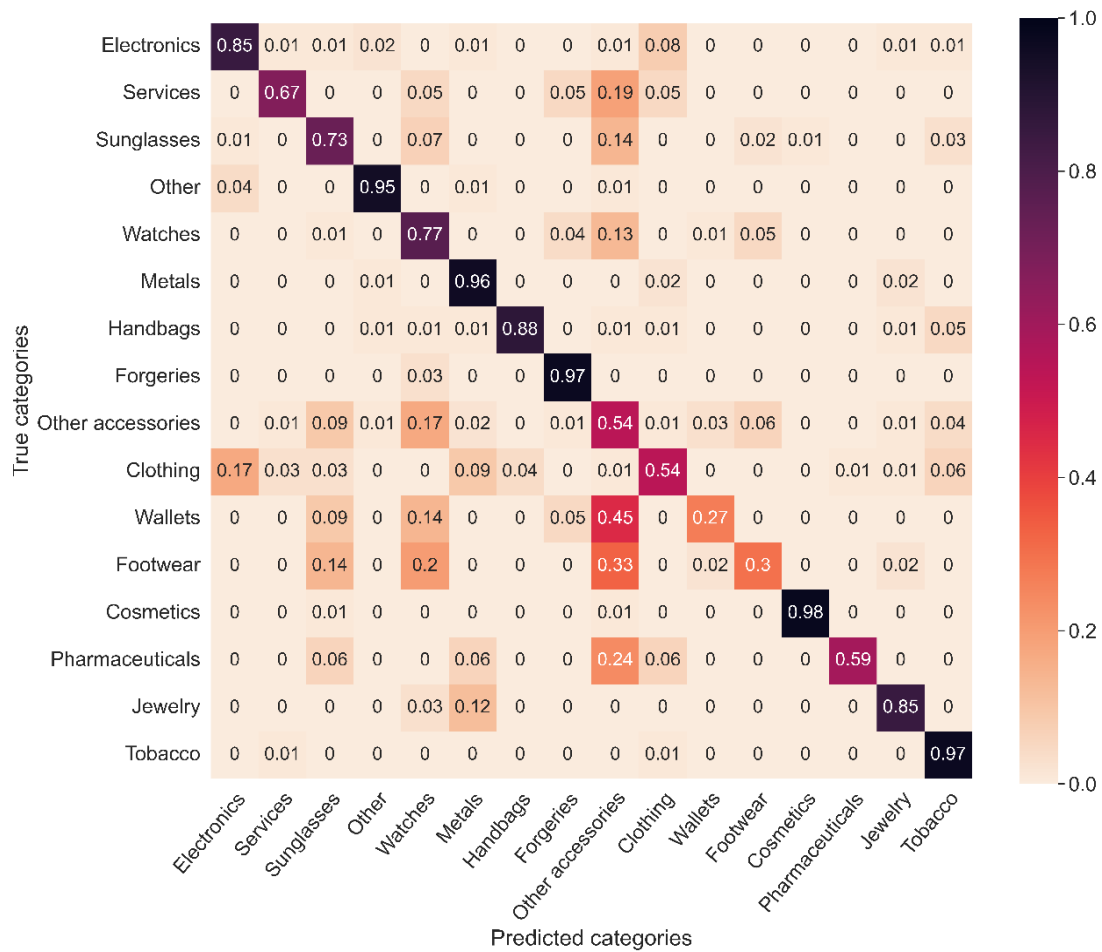

Figure 1. Normalized confusion matrix for true and predicted categories of counterfeits.

Classification performance was generally good, but we observed that six categories showed low (Cosmetics, Tobacco, Other accessories, Other) or very low (Pharmaceuticals, Services) categorization performance. Since low performances are only present with classes exhibiting few listings in the test set, most of the listings are well categorized, which is also reflected in the weighted performance scores (Tab. 3). An exception was for the category "Other", which was also less well categorized despite containing more listings than the other low-performing categories. The category "Other" often contained custom orders, with product titles such as "custom [customer name]", complicating the annotation process. Since the classifier received additional information from the product description, which was not available to the annotators, it is possible that mismatches between the annotations and product descriptions led to more misclassifications in the category "Other". For example, some custom orders might have similar descriptions as other counterfeits. Besides custom orders, the category "Other" also included guides, instructions, counterfeit art (e.g., paintings), or cars.

Having established the accuracy of the classifier to predict the unlabeled listings (i.e., label all the unannotated listings), the entire annotated data was utilized to re-train the LinearSVC classifier with the same parameters. The advantage of re-training the classifier with the entire annotated data – instead of using the best classifier from the cross-validation procedure, which is trained only on a subset of the annotated data – is that all the annotated data can be leveraged for the training, which supports better predictions.

### 2.4. Holding and placeholder prices

Previous studies about dark markets sometimes encountered holding prices, which vendors use to mark out-of-stock listings, preventing their removal from the market (Soska & Christin, 2015; van Wegberg et al., 2018). Some holding prices are very high to prevent anyone from buying the product. The advantage of a holding price is that vendors can keep showing customers what was sold and what might be coming back in stock. However, when estimating price or sale volumes on markets, holding prices with very high values can distort the actual results. Therefore, we used a heuristic proposed and used by others (Soska & Christin, 2015; van Wegberg et al., 2018) to replace high holding prices (≥ 10,000 USD) with the original price (if available) or to remove it. In addition, we also looked at listings with very low prices (≤5 USD) and found that such prices were mainly not the actual selling price and seemed to function as placeholders too. For example, many listings with a price of 0 need further specifications by the customer (often instructed in the listing description), such as amounts, colors, or shipping, which affects the final price. However, during the data scraping process, the placeholder price is mostly that which is collected rather than the individual price variations. In some instances, vendors listed the variations of the products in separate listings and later merged them into a single listing with the option of making the wanted changes (color, amount, etc.) or vice versa. In such cases, we can determine the average price of such a merged listing to get a more accurate representation of the product price. For listings with a holding and placeholder price, we searched for the same product from the same vendor to find a replacement price. Table 4 shows the distribution of found and replaced holding and placeholder prices.[8] Products with a high holding price for which we did not find replacements were excluded from further analyses of the value of the goods.

|  | n | % of all listings | Replaced | Mean price (USD) |
|---|---|---|---|---|
| **Holding (≥10,000 USD)** | 120 | 0.08 | 83 | 140.67 |
| **Placeholder (≤5 USD)** | 6,106 | 3.87 | 1,040 | 178.64 |
| **Total/Mean** | 6,226 | 3.94 | 1,123 | 159.66 |

Table 4. Number of found and replaced holding and placeholder prices and the average price of all replacements.

### 3. Results

This section looks at the data for all products and counterfeits and their distribution across markets. We then focus on counterfeit product types and product origins and compare our measures with estimates from audits of goods seized by law enforcement at borders. Lastly, we evaluate the monetary value of offered and sold counterfeits and the generated sales volume of vendors.

### 3.1. Product offers and counterfeit prevalence

Figure 2 shows how many products (not just counterfeits) were offered across all markets over time. The volumes shown are monthly and contain all available products on the dark markets. For most markets, the data range between January 2014 and September 2015, but the data for the market Alphabay extends to January 2017. Evolution and Agora offered the most products, followed by Alphabay, Abraxas, BlackBank Market, and Cloud 9. The remaining markets seem to have offered only a minimal number of products and for shorter periods. Reasons for this variation differ. For example, some markets were closed down by law enforcement (Cloud 9, Alphabay), closed down voluntarily (The Marketplace, Agora), experienced an exit scam[9] (Evolution, BlackBank Market, Andromeda, Middle Earth Marketplace, Abraxas), or were hacked (EMCDDA-Europol, 2017).[10] However, scraping data from dark markets can also be unstable, leading to gaps in the data (Ball et al., 2019; Du et al., 2018; Ghosh et al., 2017; Van Buskirk et al., 2016). Thus, we can only capture a partial picture of overall events, probably leading to underestimating the availability of products on dark markets and their value.

---
[8] If several replacement prices were available, we took the average price as the replacement.

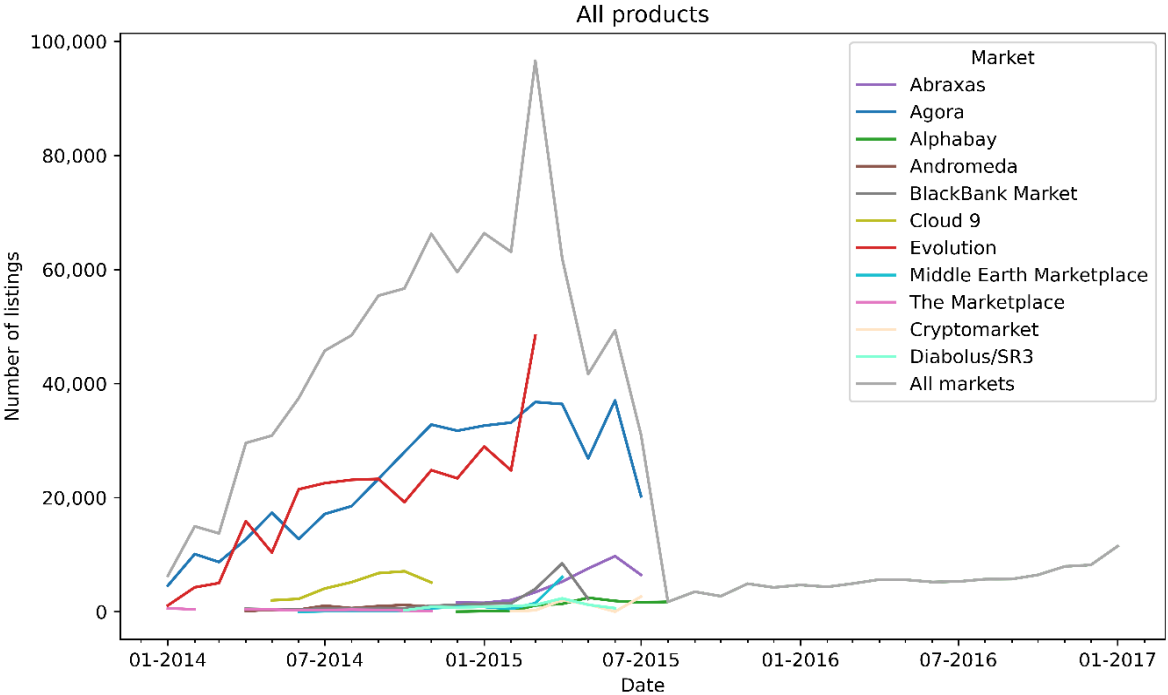

Figure 2. Monthly volume of products offered across markets.

Figure 2 also shows the monthly volume of products offered across all markets combined (gray line). Overall, product offerings seem to increase steadily, with a sharp peak at the beginning of 2015 with almost 100,000 listings. Offers then starkly declined, with only a few products on offer from mid to late 2015, followed by a slow increase for the remaining time, the latter solely attributed to Alphabay. To make comparisons and estimations of counterfeits across markets more comparable, we subsequently focus on the timeframe for which most markets had at least some listings on their platforms: January 2014 to September 2015.

Focusing on counterfeits (Fig. 3), we see a similar overall trend (gray line). However, as expected, the overall number of offers is much lower, with counterfeits accounting for around 2.69% of all listings across markets. Interestingly, the observed proportion of counterfeits on dark markets coincides well with the estimated overall proportion of counterfeits worldwide (3.3%) discussed above (OECD/EUIPO, 2019). Furthermore, only nine of the eleven markets seem to offer counterfeits, with Agora and

---

[9] An exit scam describes a situation in which the platform (market) owners steal all cryptocurrencies from all customers. On many markets it is necessary to upload cryptocurrency to an account before making a purchase. Thus, market owners have full control over the customers deposits. In some instances, the market owners also control the implemented escrow service, allowing for an even bigger exit scam.
[10] For detailed timeline of the market lifespans and their reasons for closing see (EMCDDA-Europol, 2017).

Evolution offering the most, followed by BlackBank Market, Alphabay, and Middle Earth Marketplace. The remaining markets seem to harbor only a minimal number of counterfeits. Most offers seem to occur between the beginning- and mid-2015.

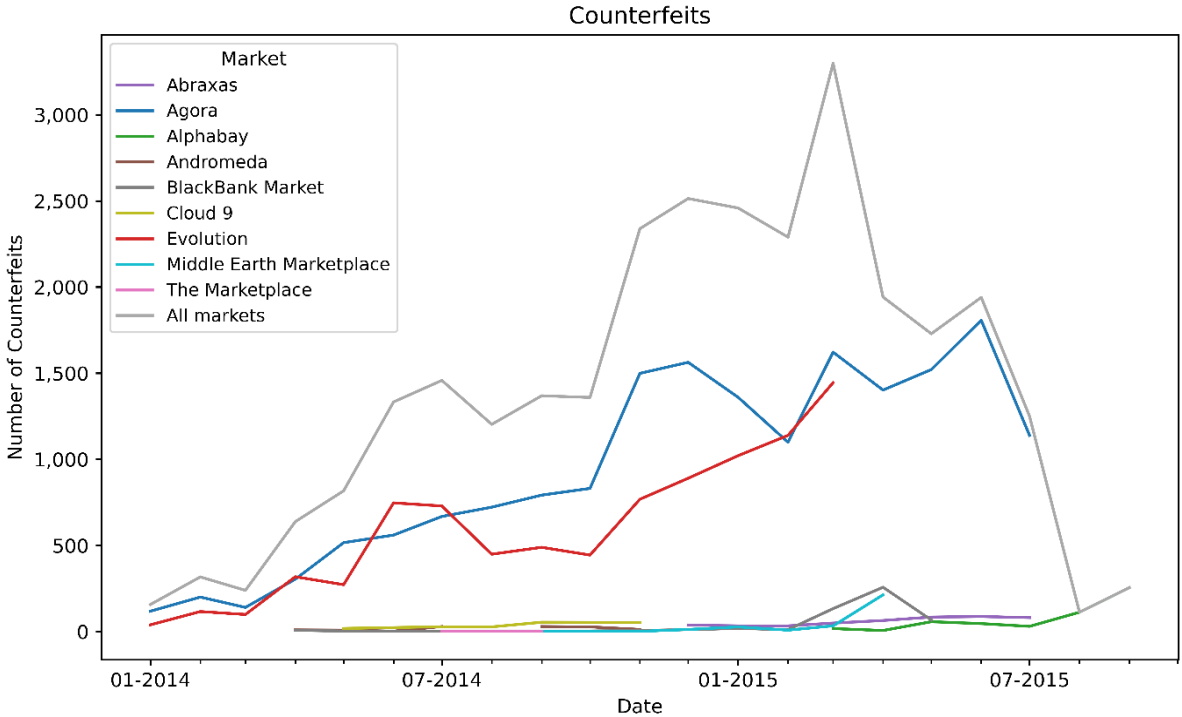

Figure 3. Monthly volume of counterfeits offered across markets.

### 3.2. Counterfeit product types and occurrences

Focusing on counterfeit product types (Tab. 5), we observe that watches make up most of all products (59%) listed on the markets, followed by four categories, each of which accounts for between 4-6% and collectively account for around 20% of all counterfeits. Most of the remaining categories contribute only a little, with most representing only 2% or less of all counterfeits. Thus, almost 80% of counterfeits listed were represented by only five (of the 16) categories of products.

By comparing our measures of the types of counterfeits to goods seized at borders, we can identify how products differ and discuss possible contributing factors to those differences. Based on a report by OECD/EUIPO (2019), which summarizes findings regarding seized counterfeits between 2014 and 2016, we see that not all categories represented on dark markets are also present in seized goods (Tab. 5). Also, the distribution of counterfeits found on dark markets and seized products varies greatly. In addition, sunglasses, handbags, and other accessories, which make up around 10% of counterfeits on dark markets, are not listed individually in the report but are grouped within headgear (1.5%), miscellaneous (0.4%), and articles of leather (13.4%). The remaining categories show a similar distribution (OECD/EUIPO, 2019).

Another report by the Intellectual Property Office (IPO) in the United Kingdom shows a different picture of IP and counterfeit-affected product categories (IP Crime Group, 2015). The report summarizes independently reported IP crimes through Crimestoppers[11] and investigations of counterfeits by Trading Standards (TS)[12] between 2014 and 2015. The top five reported and investigated IP crimes were Tobacco, optical media, clothing, alcohol, and footwear. Although watches, jewelry, cosmetics, and electronics were also within the top 17 affected categories, they seem to be less prominent than on dark markets and attracted fewer investigations by TS (Tab. 5). The differences observed for Tobacco, Footwear, Electronics, Clothing, and Watches, are further examined in the Discussion, section 4.

---

[11] Crimestoppers is a non-governmental organization, which allows citizens to anonymously report crimes and concerns (https://crimestoppers-uk.org/).

| Category | Dark Markets | OECD/EUIPO | IPO |
|---|---|---|---|
| Watches | 59.27 | 5.70 | 1.19 |
| Clothing | 5.93 | 17.7 | 7.94 |
| Other | 5.41 | - | 7.41 |
| Forgeries | 4.55 | - | - |
| Sunglasses | 4.10 | - | - |
| Electronics | 3.78 | 12.25 | 0.80 |
| Handbags | 3.42 | - | 1.44 |
| Other Acces. | 2.93 | - | - |
| Footwear | 2.92 | 22.6 | 2.77 |
| Jewelry | 2.49 | 1.85 | 0.51 |
| Services | 1.90 | - | - |
| Wallets | 1.37 | - | - |
| Metals | 0.94 | - | - |
| Pharma. | 0.49 | 1.5 | 0.30 |
| Cosmetics | 0.26 | 3.5 | 1.10 |
| Tobacco | 0.24 | - | 28.15 |

Table 5. Percentage of counterfeit categories; not all categories are shared by the reports; see appendix F for separate and complete lists of counterfeit categories by OECD/EUIPO (2019) and IP Crime Group (2015).

### 3.3. Counterfeit origins

Next, we examine the shipping origins of products as indicated on the product listings. Figure 4 shows the percentage of shipping origins for all products and counterfeits across all markets. All countries that accounted for 1% or less are aggregated into the category "Other". While possible shipping destinations are included in the listing data, we did not analyze these as most destinations are listed as "Worldwide" or "Undeclared", providing only limited information. The distribution of the shipping origins for all products seems to differ from counterfeits. However, "Undeclared" takes up a considerable portion in both cases. While most products seem to originate from the USA, most counterfeits are from China, including Hong Kong. "Other" contained mostly European countries (e.g., Italy, France, Poland, Portugal), it also contained a range of Asian countries (India, Thailand, Singapore, Cambodia), and others (e.g., Afghanistan, Chile). The category "EU" (Europe) is not an aggregation we generated but was indicated on some products. Thus, for those products, we cannot say which European countries they originate from specifically.

---

[12] Trading Standards is the local law enforcement within the UK, investigating IP crimes and enforce consumer protection legislations (https://www.tradingstandards.uk/).

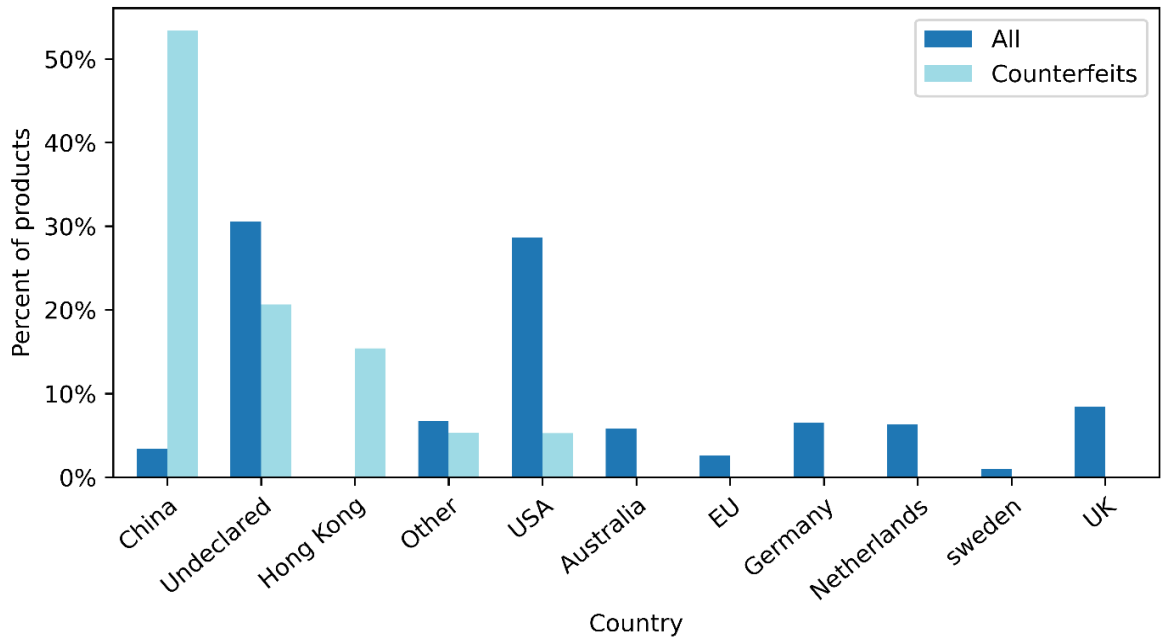

Figure 4. Percentage of shipping origins for all products and counterfeits.

Table 6 shows the association between particular types of counterfeit goods and the country they were listed as originating from. Each row shows where listings for the product category originated. Countries that accounted for less than 10% of the listings were aggregated into the category "Other". For example, 74.92% of counterfeits categorized as footwear originated from China and 25.08% from "Other". The countries of origin are mutually exclusive, and so the row totals sum to 100%. As previously indicated, China is well represented, contributing to many categories. For Cosmetics, Electronics, Pharmaceuticals, and Services, additional countries previously included in the "Other" category are now visible and seem to specialize in supplying one particular type of counterfeit. However, for some counterfeits, the "Other" category accounts for a substantial fraction of counterfeits indicating that in these cases, the products originate from a large number of countries. In addition, only six categories (Footwear, Clothing, Cosmetics, Pharma., Tobacco, and Watches) seem to have a rate of undeclared origins of below 20%, possibly indicating that many sellers are concerned about giving up too much information by indicating a product origin. Since Belgium is a relatively small country, we examined the services that seemed to originate from there more closely. Of the 195 unique offered services for all counterfeits, 11.86% originated from Belgium (BE), representing 23 listings. All services originating from Belgium were listed by a single vendor and included digital goods, such as Facebook likes, guides about making money, using Tor, and hacking. Since digital goods are not reliant on physical transportation, their origin might only inform us about the possible residence of the individual offering such goods. We also manually examined the products sold by Austria (AT), Australia (AU), Thailand (TH), Afghanistan (AF) and Germany, as they each sold items of one product category that accounted for at least 10% of the total. Australia and Germany both contribute to the sales of pharmaceuticals. Two vendors from Australia contributed nine listings to pharmaceuticals, selling drugs (that were misclassified), bust also fake drugs (presumably used for dilutions to increase profit margins), and measuring syringes. Three vendors from Germany contributed five pharmaceutical listings, which were misclassified chemical drugs, showing that product descriptions from drug listings and pharmaceutical counterfeits can be very similar. Two vendors from Thailand offered 48 electronics, mostly counterfeited smartphones and headphones. 41 handbags that originate from Afghanistan were mostly imitations of Louis Vuitton handbags. 16 cosmetics originated from Austria: counterfeit perfumes from Chanel and Luca Bossi.

| Product category | China | Hong Kong | USA | AT | AU | TH | AF | BE | DE | EU | Other | Undc. |
|---|---|---|---|---|---|---|---|---|---|---|---|---|
| Footwear | 74.92 | 0.00 | 0.00 | 0.00 | 0.00 | 0.00 | 0.00 | 0.00 | 0.00 | 0.00 | 25.08 | 0.00 |
| Watches | 69.75 | 22.01 | 0.00 | 0.00 | 0.00 | 0.00 | 0.00 | 0.00 | 0.00 | 0.00 | 8.23 | 0.00 |
| Clothing | 63.70 | 0.00 | 0.00 | 0.00 | 0.00 | 0.00 | 0.00 | 0.00 | 0.00 | 0.00 | 26.24 | 10.07 |
| Jewelry | 48.63 | 0.00 | 0.00 | 0.00 | 0.00 | 0.00 | 0.00 | 0.00 | 0.00 | 0.00 | 19.61 | 31.76 |
| Sunglasses | 40.81 | 0.00 | 17.90 | 0.00 | 0.00 | 0.00 | 0.00 | 0.00 | 0.00 | 0.00 | 20.76 | 20.53 |
| Other Ac. | 36.33 | 11.67 | 13.33 | 0.00 | 0.00 | 0.00 | 0.00 | 0.00 | 0.00 | 0.00 | 1.00 | 37.67 |
| Electronics | 23.06 | 0.00 | 25.39 | 0.00 | 0.00 | 12.44 | 0.00 | 0.00 | 0.00 | 0.00 | 14.25 | 24.87 |
| Wallets | 20.00 | 23.57 | 0.00 | 0.00 | 0.00 | 0.00 | 0.00 | 0.00 | 0.00 | 0.00 | 8.57 | 47.86 |
| Handbags | 11.71 | 0.00 | 14.00 | 0.00 | 0.00 | 0.00 | 11.71 | 0.00 | 0.00 | 0.00 | 10.29 | 52.29 |
| Tobacco | 0.00 | 0.00 | 76.00 | 0.00 | 0.00 | 0.00 | 0.00 | 0.00 | 0.00 | 0.00 | 24.00 | 0.00 |
| Pharma. | 0.00 | 0.00 | 54.00 | 0.00 | 18.00 | 0.00 | 0.00 | 0.00 | 10.00 | 0.00 | 18.00 | 0.00 |
| Metals | 0.00 | 0.00 | 35.42 | 0.00 | 0.00 | 0.00 | 0.00 | 0.00 | 0.00 | 0.00 | 16.67 | 47.92 |
| Other | 0.00 | 0.00 | 15.01 | 0.00 | 0.00 | 0.00 | 0.00 | 0.00 | 0.00 | 0.00 | 24.95 | 60.04 |
| Services | 0.00 | 0.00 | 12.37 | 0.00 | 0.00 | 0.00 | 0.00 | 11.86 | 0.00 | 0.00 | 7.73 | 68.04 |
| Forgeries | 0.00 | 0.00 | 12.26 | 0.00 | 0.00 | 0.00 | 0.00 | 0.00 | 0.00 | 0.00 | 25.16 | 62.58 |
| Cosmetics | 0.00 | 0.00 | 0.00 | 59.26 | 0.00 | 0.00 | 0.00 | 0.00 | 0.00 | 11.11 | 14.81 | 14.81 |

Table 6. Percentage of counterfeit shipping origins by country and product category; percentages are split by countries and aggregate to 100% for each row; AT = Austria; AUS= Australia; TH = Thailand; AF = Afghanistan; BE = Belgium; DE = Germany; EU = Europe; Undc. = Undeclared.

In contrast to the differences observed for counterfeit products seized at borders and offered on dark markets, product origins seem to match better across data sources. For example, between 2014-2016, seized goods mainly originated from China (55%) and Hong Kong (26.2%) (EUIPO, 2019; OECD/EUIPO, 2019). However, seized goods also originated from the United Arab Emirates (3.8%), Turkey (3.1%), Singapore (2.8%), Thailand (1.4%), India (1%), and other countries (each with less than 1%) (OECD/EUIPO, 2019). In contrast, for the dark markets, counterfeits were either not explicitly offered from these countries (e.g., Singapore, Thailand, India), or they accounted for less than 1% of the listings. Interestingly, the USA seems to account for 5% of counterfeits on dark markets while only accounting for 0.4% in seized goods.

### 3.4. Counterfeit prices, sales volume, and surface web prices

Lastly, we summarized counterfeit prices for each category (Tab. 7), estimated vendor sales volumes (Tab. 8 and Fig. 5), and examined the price differences of products offered on darknet markets and the surface web (Tab. 9, Fig. 6).

#### 3.4.1. Observed counterfeit prices

Table 7 shows the prices for all counterfeit listings (offers) as customers can see them on the markets. Prices are expressed in USD and are based on all counterfeit listings at the time the listing was posted.[13]

---

[13] All listings which contained only cryptocurrency prices were transformed into USD, utilizing the average conversion value from "https://coinmarketcap.com/currencies/bitcoin/historical-data/" on the day the listing was dated (scraping date).

The total price volume represents the accumulation of all prices from all unique counterfeits for each category (i.e., the total value if each listed item would have sold once, but only once). The total price volume of all unique counterfeits from Jan-2014 to Sep-2015 is around 1.8 million USD. Many maximum prices of each counterfeit category are high, often attributed to wholesales. The highest observed mean price is for metals, including collectible gold and silver coins or bullions, while the lowest is for sunglasses. With watches making up most listings, they also hold the highest volume, around 1 million USD. Minimum prices of 0.00 are mostly placeholders, and are not free products, often used to prompt the user to select an amount, color, model, and so on (see above). Both Metals and Pharmaceuticals show high standard deviations, which can be attributed to a few very high-priced listings. For example, "28g PSEUDO SPEED" for $2000 or "Lot of 10 High Quality Counterfeit Gold Bars" for $5799.

The last two rows show the prices' mean, total, and weighted mean. Specifically, in the "Mean/Total" row, each USD column (Min, Max, Median, etc.) is averaged by dividing the sum of all product category prices by the number of product types, while solely the column "# Listings" is totaled. The weighted mean is the result of taking the sum of the product of the category price and the number of listings of the same category, divided by the total number of listings. Thus, each mean is weighted by the number of listings available in each product category.

|  | Price in USD ($) | | | | | | |
|---|---|---|---|---|---|---|---|
| **Category** | **Min** | **Max** | **Median** | **Mean** | **SD** | **Volume** | **# Listings** |
| Tobacco | 0.00 | 1,401.25 | 0.00 | 110.82 | 315.07 | 2,770.52 | 25 |
| Cosmetics | 13.50 | 1,512.11 | 27.91 | 191.51 | 382.34 | 5,170.71 | 27 |
| Wallets | 0.00 | 285.04 | 88.69 | 100.69 | 61.86 | 14,096.17 | 140 |
| Jewelry | 0.00 | 856.27 | 38.34 | 57.55 | 114.00 | 14,503.22 | 255 |
| Sunglasses | 0.00 | 174.48 | 43.16 | 44.35 | 21.23 | 18,536.27 | 419 |
| Pharma. | 0.10 | 9,869.19 | 50.52 | 421.12 | 1,443.72 | 20,634.70 | 50 |
| Footwear | 0.00 | 310.01 | 74.22 | 90.46 | 45.82 | 26,322.84 | 299 |
| Services | 0.00 | 4,901.10 | 29.66 | 225.90 | 580.18 | 43,824.74 | 194 |
| Handbags | 0.00 | 1,570.34 | 96.04 | 127.10 | 145.00 | 44,358.32 | 350 |
| Metals | 0.00 | 5,413.99 | 161.68 | 524.47 | 1,029.88 | 49,824.93 | 96 |
| Clothing | 0.00 | 9,878.84 | 30.00 | 87.06 | 409.30 | 52,669.08 | 606 |
| Electronics | 0.00 | 2,988.94 | 84.96 | 165.07 | 271.64 | 63,220.44 | 386 |
| Other Acc. | 0.00 | 1,530.65 | 68.33 | 253.03 | 423.32 | 73,125.05 | 300 |
| Forgeries | 0.00 | 7,291.15 | 48.77 | 286.19 | 722.87 | 129,931.09 | 465 |
| Other | 0.00 | 5,516.25 | 118.51 | 321.20 | 675.91 | 225,805.96 | 553 |
| Watches | 0.00 | 3,957.15 | 100.91 | 171.03 | 178.79 | 1,017,309.52 | 6,060 |
| **Mean/Total** | 0.85 | 3,591.05 | 66.36 | 198.60 | 426.31 | 112,631.47 | 10,225 |
| **Weighted mean** | 0.04 | 4,013.13 | 87.22 | 174.24 | 262.70 | 633,721.53 | - |

Table 7. Summary counterfeit prices and volumes for each product category in USD; Mean = column mean prices.

### 3.4.2. Estimated counterfeit sales volumes

As in previous research (Soska & Christin, 2015; van Wegberg et al., 2018), we utilized the total number of feedback comments provided for each listing to estimate how often an item was sold. Since the data was scraped recurrently, listings and their associated feedback is collected cumulatively, adding old and new feedback for every scrape completed. To avoid duplication, we only analysed unique items of feedback. The number of unique feedback was then multiplied by the product's listing price on the darknet market ($PP_{DM}$) to obtain an estimated sales volume in USD (Table 8). Although some markets made feedback mandatory in the past, we do not know how the markets in this analysis regulated feedback and if fake reviews are moderated, resulting in some uncertainty as to whether the review count is an under- or over-estimation of sales. However, based on the current estimates, most sales were for watches, followed by "Other" (6.50%) and Forgeries (5.96%).

| Category | Estimated Sales Volume | | Total Feedback |
|---|---|---|---|
| | Total USD | Share (%) | |
| Cosmetics | 16.98 | 0.01 | 1 |
| Tobacco | 52.90 | 0.04 | 2 |
| Pharma. | 579.60 | 0.45 | 20 |
| Metals | 675.71 | 0.53 | 5 |
| Wallets | 678.17 | 0.53 | 9 |
| Footwear | 1,250.46 | 0.98 | 16 |
| Jewelry | 1,464.83 | 1.15 | 31 |
| Services | 1,515.03 | 1.19 | 51 |
| Other Acc. | 1,948.04 | 1.53 | 40 |
| Handbags | 2,809.64 | 2.20 | 35 |
| Clothing | 2,936.27 | 2.30 | 35 |
| Sunglasses | 3,356.94 | 2.63 | 84 |
| Electronics | 7,176.75 | 5.63 | 55 |
| Forgeries | 7,604.31 | 5.96 | 31 |
| Other | 8,288.46 | 6.50 | 78 |
| Watches | 87,160.41 | 68.35 | 439 |
| **Total** | **127,514.50** | **100** | **932** |
| **Calculation** | $\sum PP_{DM} \cdot F$ | $\frac{SV}{Total\ SV} \cdot 100$ | Count |

Table 8. Estimated sales volume (USD) for each category based on the number of feedbacks; $PP_{DM}$ = Product Price on the dark market; F = Number of feedback comments; SV = Sales volume.

Considering the monthly sales volume by category (Fig. 5), we observed a similar trend for available listings (Fig. 3), with most sales occurring between mid-2014 and mid-2015. We also observed two peaks in sales in mid-2014 and mid-2015. Again, watches are represented most, followed by forgeries and "Other".

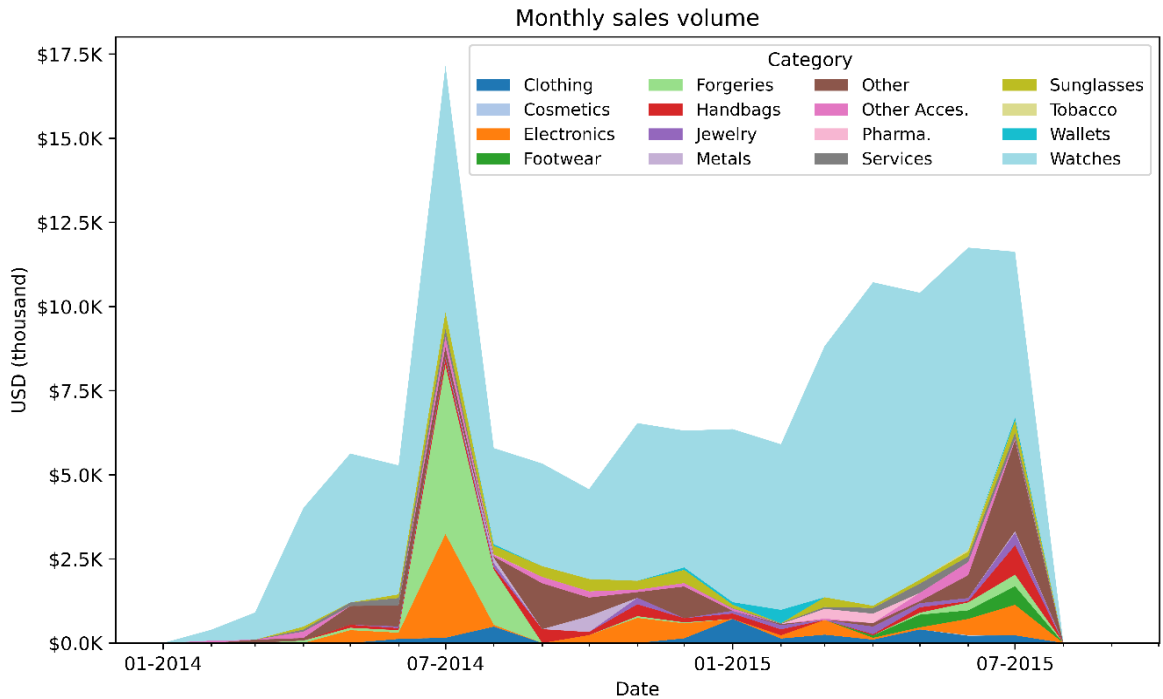

Figure 5. Monthly stacked sales volume (based on feedback) across categories.

Comparing these figures to seizures at borders, a report by the OECD/EUIPO (2019) found that the largest value share for goods seized at borders was for watches (22.9%), followed by leather articles (11.6%), electrical equipment & machinery (10.8%), footwear (10.5%), clothing (8.2%), jewelry (5.9%), cosmetics (4.9%), toys (4.6%), optical/photographic & medical instruments (4.1%), mechanical appliances (1.5%), vehicles (1.4%), and other products (less than 1%). Although watches seem to account for the most value in dark markets *and* border seizures, the concentration of watches is much more pronounced for the dark markets, with an estimated sales volume of over 68%. Thus, seized goods appear to show a more equally distributed range of values across products than is observed on the darknet markets. Categories, such as machinery, toys, medical instruments, and appliances listed for seized goods, did not appear to be explicitly sold on dark markets, probably contributing to the skewed product distribution observed there.[14]

### 3.4.3. Dark and surface market prices

In addition, we sampled ten darknet market products from each category and determined their price on the surface web (Tab. 9). For 25 products, we determined the historical price on the surface web by utilizing a product price comparison site (geizhals.eu)[15] which records the complete price development over the product's lifespan.[16] We determined the current price using Google Shopping (shopping.google.com) for the remaining products.[17] When possible, we used the prices from original brand stores (e.g., Hermes, Louis Vuitton, Gucci, etc.) but selected prices from other shopping platforms if the products were not manufactured anymore or were not otherwise listed. For 46 dark market products, we found the exact match on the surface web, while for the remaining listings, we selected the next best match from the same brand.[18] For the category Metals, we adjust different indicated weights on the listings (e.g., 10 ounces, 1 gram, 1Kg) by extrapolating the cost for 1 ounce for each listing, making a comparison possible. We excluded products from the categories Services, Forgeries, Pharmaceuticals, and "Other" since most of these products cannot be purchased on the surface web.[19] Product prices in Euro were converted into USD based on the conversion rate present on the price date. Since we selected only ten random samples for each product category, the estimated price differences are only intended to illustrate the observed trend and should not be regarded as a complete analysis.

---

[14] The category "Other" contained a high variation of different products, including leather products (e.g., belts) and cars.
[15] Geizhals.eu collects prices from original, licensed vendors, and reselling platforms (e.g., eBay), capturing price developments after the products is not manufactured anymore.
[16] We determined the historic price by looking up the price of the product on the date it was listed on the dark market. For four products, we found historic prices deviating from the listing date by 2-4 months.
[17] Google shopping shows products from original, licensed vendors, and reselling platforms (e.g., eBay).

| Category | Mean [SD] |
|---|---|
| Cosmetics | 112.82 [79.94] |
| Sunglasses | 195.40 [116.98] |
| Tobacco | 331.21 [433.75] |
| Electronics | 386.07 [448.00] |
| Wallets | 867.22 [966.92] |
| Metals (1oz.) | 1,084.08 [755.42] |
| Other access. | 1,167.03 [1,654.88] |
| Footwear | 1,263.27 [2,767.35] |
| Handbags | 1,351.87 [856.22] |
| Jewelry | 1,712.02 [1,318.34] |
| Clothing | 2,391.83 [6,712.34] |
| Watches | 25,338.95 [52,631.70] |
| **Average** | **3,464.48 [6733.06]** |

Table 9. Mean [SD] of 10 sample products for each category on surface markets.

To better understand the relationship between darknet markets and surface web prices, we plot one against the other in Figure 6. Across all product categories, products are more expensive on the surface web, but prices between and within categories vary considerably. Prices between darknet markets and the surface web are closest for cosmetics (for which the mean ratio was 2.22) and most different for watches, which were, on average, 147.23 times more expensive on the open than the dark web.

---

[18] In some cases, the product titles from the dark markets were not detailed enough (e.g. "LV wallet") to find the exact products on the surface web. Three products in cosmetics and five in Tobacco could not be found on the surface web.
[19] While randomly sampling products from "Other" we encountered products such as a car code grabber, a mail list for spam, or other digital services, which are not sold on the surface web. We also encountered a counterfeit of a Picasso painting "Seated Woman (Marie-Therese)", worth more than USD 60 million.

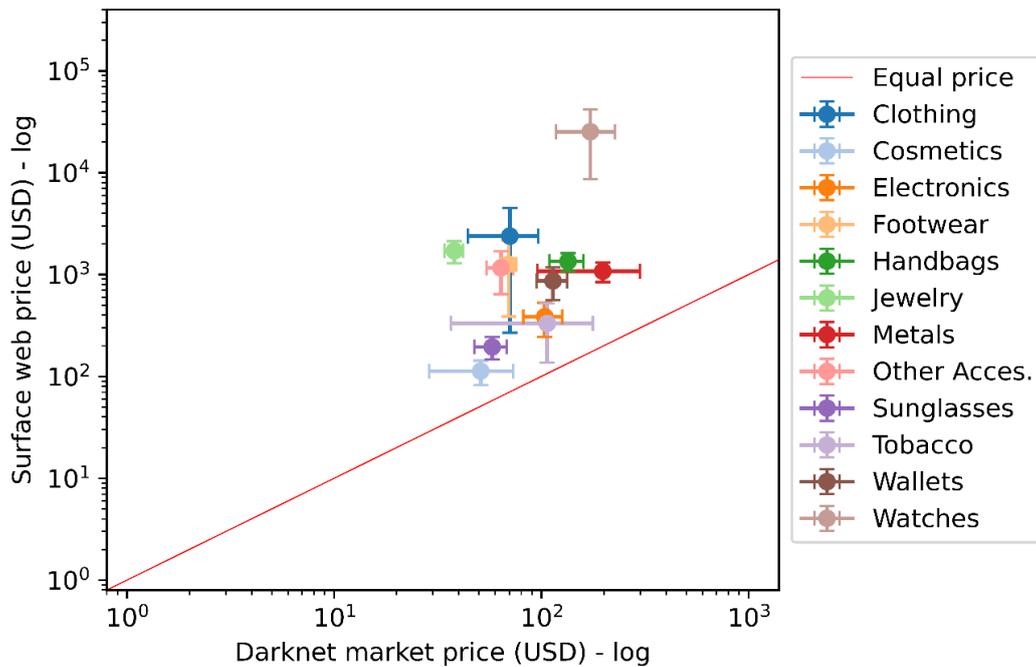

Figure 6. Mean (SE) price differences for each product category between Darknet markets and the surface web (See appendix G for price differences of each product).

## 4. Discussion

Insights about counterfeits typically originate from data on goods seized at borders by law enforcement agencies. As discussed, these data are not collected through random sampling or other approaches that would ensure that the findings are representative of the ground truth. Instead, they are subject to various biases associated with the intelligence that law enforcement agencies collect or have access to, or the policies followed at borders. This means that our understanding of what is counterfeited is likely to be biased. Further indications of possible biases can be found in the prevalence estimation differences for various agencies (IP Crime Group, 2015; OECD/EUIPO, 2019). Given that IP crime is known to be increasing (Federal Bureau of Investigation, 2014, 2015, 2016; OECD/EUIPO, 2019), it is important to understand the counterfeit economy better, and in this study, we examine what insights the analysis of data regarding the availability and price of counterfeits on dark markets might provide.

### 4.1. Product categories

The current study suggests that the share of counterfeits on dark markets (2.69%) seems to be slightly above previous expectations, which were around 1.5-2.5% (Europol, 2017). We also see differences in some product categories observed during seizures and counterfeits offered on dark markets. As already described, seized products are most likely biased through the activities and procedures adopted by authorities affecting estimations on which product types are affected. Examining the counterfeit categories, we see that watches account for most of the value in both cases but are more prominent on dark markets overall. Watches might be more challenging to identify or detect as counterfeits as other products (e.g., shoes, clothes, Tobacco) in seizures, perhaps due to very high-profit margins, an increased effort is put into making fake watches more difficult to identify. Alternatively, watches might be less prone to bulk shipments and make their way through borders differently than other items (e.g., single parcel shipments through the air versus containers at ports). Hence, watches might be shipped more diversely, possibly going through different security measures and being more difficult to catch overall. However, single parcel shipments might only be worthwhile for high-value items, such as watches, but less profitable for items that need high-volume sales.

Interesting are also other strong prevalence estimation differences, such as for Tobacco, Footwear, Electronics, and Clothing. For some product groups, estimations from authorities are missing entirely (e.g., sunglasses, handbags, accessories, wallets, metals, Tobacco). Especially, Tobacco seems to make up only 0.24% of counterfeits on dark markets, which is missing in estimations by OECD/EUIPO (2019) but is highly representative in measures by IP Crime Group (2015). Vendors on the dark market might favor high-value products, possibly tailoring more towards end-consumers than other businesses. Thus, Tobacco might be more difficult to sell in high volumes on darknet markets. Similarly, OECD/EUIPO (2019) measured relatively high ratios of Footwear, Clothing, and Electronics, which are far less prevalent on the dark market. Again, such differences might originate from biases in selecting shipments for inspections but also illustrate the current issue of inconsistent measurements capturing what is being counterfeited. Important to note is that the authorities also seized counterfeits that are missing on darknet markets, such as vehicles, furniture, or alcohol, which can distort the ratio of product groups (see Appendix F for a full list of seized goods).

### 4.2. Product origins

Seized and dark market counterfeits mostly seem to originate from China and Hong Kong. However, some uncertainty surrounds the information about the origins of dark market counterfeits since providing this information is voluntary, and a large portion is undeclared (see Limitations). Nonetheless, the stark outlier in product origins of seized goods and product offers on dark markets is the US. Around 5% of dark market counterfeits were listed as originating from the USA, while only 0.4% of goods seized at borders come from the US. Again, such a discrepancy might be due to biased expectations by law enforcement, as searches are sometimes based on shipment origins (Männistö et al., 2021). Thus, border seizures might miss counterfeits originating from countries suggested by dark markets, such as the US. For example, Tobacco, pharmaceuticals, metals, electronics, and accessories (e.g., sunglasses) could be scanned for counterfeits when originating from the US. Similarly, cosmetics seem to originate from Austria more frequently, and pharmaceuticals from Australia. Alternatively, counterfeits from the US might be more heavily purchased domestically, leading to limited exportation, which would avoid border controls. Moreover, dark market listings represent the availability of a product rather than the actual supply of them. Although knowing which country counterfeits are available is helpful, products must be purchased first and subsequently shipped to be found at a border. Thus, estimation of product origins from dark markets and measures of seized goods might also vary because they capture products at different supply chain stages.

### 4.3. Vendor sales volume and product values

Similarly, estimating the sales volume and monetary value of counterfeits on dark markets is accompanied by uncertainty, which is further addressed in the next section (Limitations). However, we can see that the estimated sales volume generated for counterfeits on dark markets seems very small compared to the possible value of the items on the surface web. Europol (2017) estimated that most physical counterfeits on dark markets are sold for one-third of the actual price. Based on the current study, the discrepancy between counterfeit prices and their actual values on the surface web are more diverse and can be twenty times larger (e.g., for watches). Such differences suggest that the prices and possible sales volumes depend highly on the product category. However, the current price differences illustrate that purchasing darknet market counterfeits and selling them on the surface web could lead to considerable profits. Thus, it might be helpful to focus the attention of authorities on highly valuable counterfeits, such as watches, clothing, or jewelry, as they seem to generate the biggest profits. Notably, relative to the patterns observed for darknet markets, watches were underrepresented in the estimates based on seizures, and metals were not featured at all.

We can also see greater differences between dark and surface web prices for higher-value products, such as watches, clothes, and jewelry. Dark market vendors might prioritize higher-valued products, which can generate profits faster than products with lower profit margins (e.g., accessories, Tobacco). Such a strategy would support the idea that darknet market vendors might tailor their products more towards end-consumers, who purchase fewer items, rather than businesses, which could purchase items in high volumes with the purpose of re-selling them. In other words, lower profit margin products need higher turnovers for high profits, which is facilitated by business-to-business transactions.

### 4.4. Possible preventative measures

Since the darknet market counterfeits identified here were fully manufactured consumer products, for them to enter the supply chains of legitimate retailers, the latter would either have to sell them knowingly, or they would have to be introduced during distributional processes so that they are mixed with genuine products without the retailer's knowledge. Counterfeits could be introduced during packaging, distribution to wholesalers, retailers, or any other transportation process. As Hollis & Wilson (2014) discuss, addressing the problem in cases where companies have been misled would involve improvements to guardianship in risky parts of the supply chain. Companies could be provided with information about which products are affected and from which country they originate to facilitate their efforts to identify risks in their supply chain. For example, informing personnel that are responsible for overseeing the distribution of an affected product (which is being counterfeited) could help them to implement or re-evaluate their internal working processes to reduce the risk of counterfeits entering their supply chain and increasing the risk of discovery for the counterfeiters (Hollis & Wilson, 2014). Such implementations could include raising employee awareness of the affected products, implementing reporting mechanisms, or introducing additional validation checks for particular product types for specified periods of time. To aid in this activity, dark net market data – searchable by brand – could be made accessible to companies. Since product information is quite detailed, an implementation with up-to-date darknet market data is feasible.

Another issue concerns the leaking of product designs. One approach to help address this would involve the identification of products that are found to be offered on darknet markets before their official release on the surface web. Knowing that plans were shared would help companies narrow down which processes would have to be reviewed and where measures should be put in place to ensure adequate guardianship. Such measures might involve limiting access to project plans to only those who need to know about them (to minimise insider threats) and ensuring that all data are secure (to minimise external threats). While some cyber security and brand protection organizations advertise dark web monitoring to detect data leakages, such as personal data, to what extent they track counterfeits is unclear (Corsearch, 2023; Lenaerts-Bergmans, 2023).

Other approaches to counterfeiting might involve one or more of the 25 techniques of situational crime prevention (Clarke, 1995; Freilich & Newman, 2018), which are also informed by Rational Choice theory and the RAA. One such technique is target hardening, which aims to make the target of an offence (e.g. counterfeiting a product) less viable for the offender. Knowing which counterfeits are offered on darknet markets could help companies to make those products more difficult to counterfeit. For example, companies could change the used materials or manufacturing process to increase the efforts of imitating the product. Traceability of genuine products within a supply chain would also fall within that category, as it increases the efforts needed to counterfeit them, which could be technologically facilitated (Gayialis et al., 2022). Alternatively, the offenders' rationalisation for committing a crime could be challenged by removing possible excuses for their actions. Removing excuses includes approaches such as setting up rules or posting instructions to reduce ambiguity in situations that can be exploited. Such strategies could be helpful to deter employees in situations in which they could act maliciously (e.g., stealing plans, reintroducing counterfeits, sharing manufacturing or packaging plans, etc.) by reminding them what actions are disallowed or how specific work tasks should be performed (Freilich & Newman, 2018).

### 4.5. Future studies

Given the results of this study, it would be interesting to examine if and how such information about counterfeits on dark markets can be utilized as intelligence for law enforcement activities, companies, or policymakers. Since darknet markets and single vendor shops are continuously growing, data concerning counterfeit listings are likely to increase too (Labrador & Pastrana, 2022; Laferrière & Décary-Hétu, 2022; Platzer et al., 2022). Thus, the monitoring of darknet market counterfeits might also be increasingly valuable. Besides validating findings from seized goods, dark markets could serve as indicators of early trends for the onset of activities on the surface web. For example, future work could establish a monitoring system that collects dark market counterfeits data. Such a dataset could be used to search for dark market products on the surface web (e.g., Amazon, eBay) to establish if the same or similar products are sold across platforms. Furthermore, a longitudinal study could explore temporal trends, and examine if products tend to appear first on the dark web and subsequently find their way to the surface web. A common problem with research concerning dark markets is the accurate estimation of sales value. While utilizing customer feedback to make informed estimations is helpful, future work could explore if it is possible to exploit transactional data associated with cryptocurrencies used by dark markets (ElBahrawy et al., 2020; Nadini et al., 2021) to complement sales assessments of specific products.

### 5. Limitations

The data analysed here misses some bigger markets, such as the first Silkroad, Hydra, Empire, Hansa, Wall Street, and Sheep. The reasons for their exclusion were that they were not included in the data archive or lacked sufficient product categorization needed for the current analyses. The data also does not cover possible user-to-user transactions, which bypass the markets altogether (Nadini et al., 2021). Thus, the findings reported here do not reflect the entire dark market economy, just the activity recorded for those markets sampled. Furthermore, the present analyses utilized historical data without newer scrapes (see ElBahrawy et al., 2020), limiting some of the possible contemporary policy and prevention implications. However, previous work has not provided us with an understanding of how extensive counterfeits are present on darknet markets and re-using existing data in the current study serves as a proof of concept, showing that darknet market data can be valuable in understanding the counterfeit economy better.

Furthermore, a general problem related to research with dark markets is that the data collection procedure is constrained by the scraping process and individual platform closures, which can lead to gaps in the available data, which can make exact measurements of dark market activity difficult. The scraping process can be disrupted due to the slow connection of the Tor network, security measures of the website that are implemented to hinder automated data collection (e.g., required log-ins due to set session time-outs or solving recurring captchas), or temporary website closures. Thus, the overall number of observed listings and associated estimations will be more uncertain, making general conclusions more difficult.

While comparing seized counterfeits to dark markets counterfeits can help us understand how the two areas relate to each other, the comparison is only partly applicable. Dark market listings are offers, while seized products may already have been sold. Although seized products can also inform us about offers, they are only a subset of sold counterfeits from the overall market. Thus, comparisons of dark market listings with seized goods are informative, but they do not always encompass the same measures.

Similarly, uncertainties are present with shipping information and feedback associated with dark market listings. It is voluntary for vendors to make product origin declarations, and many choose not to do so. Nonetheless, many declared origins are in line with the origins of seized goods, providing us with some confidence in our measures. Also, information on postage times and possible tracking numbers is highly valued amongst customers, often referred to in feedback, making a genuine declaration of origin more attractive to vendors. Therefore, we cannot say how accurate product origin declarations are, but some incentives exist for vendors to make truthful indications.

As for product feedback, we cannot always know whether they are mandatory and whether the feedback is for a single or bulk purchase. Thus, the calculated sale volumes are approximations and will come with a general uncertainty because not all purchases will have produced feedback, one instance of feedback might be counted as a single purchase, or feedback could be artificially created to generate trust (Dellarocas, 2006). Similarly, our value estimation process should be taken with caution. Taking ten random samples for each product category will produce only rough estimates and was only intended to illustrate the estimated difference between prices on darknet markets and the surface web. Furthermore, a historic price could not be obtained for all product samples, and prices can vary considerably over time (e.g., original soccer shirts or Nike shoes), influencing estimations.

## 6. Conclusion

Based on the analyzed darknet market data, we can say that counterfeit goods are rare (2.69% of all products) on dark markets and are often included in miscellaneous categories. Thus, accurately measuring the prevalence of counterfeits across the dark web is difficult. However, we disentangled product categories using a classification model, allowing for a more in-depth analysis. We showed that some product types exhibit a strong prevalence discrepancy between dark markets and seized goods. Specifically, watches are more prominent on dark markets, while electronics, shoes, clothes, and Tobacco are more prevalent among seized goods. Furthermore, vendors seem to favor high-value products with big profit margins (e.g., watches) instead of products for which higher turnovers are necessary (e.g., Tobacco) to obtain the same revenues. Interestingly, we found some similarities in shipping origins between dark markets and seized goods, with some exceptions, such as relatively high origin shares from the US in dark market counterfeits.

While the study is based on historical data, we showed that examining dark market counterfeits in more detail can contribute to our understanding of the counterfeit market. With an increasing emergence of darknet markets and single vendor shops, offers of counterfeits are also likely to increase. Thus, examining current dark market data would be valuable in future analyses of IP crime, which would provide us with more up-to-date insights. Collecting data from dark markets to gather intelligence could be done manually and automatically and would probably be very cost-effective compared to (border) seizures. Once implemented, prolonged data collection could be easily maintained, providing us with regular details on counterfeits. Such information would be usable by authorities and businesses, informing them which products are currently affected.

# Appendix

## Appendix A: List of considered markets

Cell properties: Yellow: Not in the archive; Orange: timeframe too short or data gaps too big; Blue: Products are not or only partly categorized; Grey: Data gaps too big or no counterfeits were found within the categories; Green: market included in analyses

| Name | Archived | Data-start | Data-end | Data Categorized? | Notes |
|---|---|---|---|---|---|
| Abraxas | Yes | 16.12.2014 | 05.07.2015 | Yes | |
| Acropolis | No | - | - | - | |
| Agora | Yes | 01.01.2014 | 07.07.2015 | Yes | |
| Alpaca | Yes | 24.04.2014 | 07.11.2014 | Partly | |
| AlphaBay | Yes | 21.12.2014 | 28.01.2017 | Yes | |
| Anarchia | Yes | 07.05.2015 | 05.07.2015 | Partly | |
| Andromeda | Yes | 12.04.2014 | 18.11.2014 | Yes | Strong variation on captured listings per scrape |
| Apple Market | No | - | - | - | |
| Area51 | Yes | 22.06.2014 | 23.01.2015 | No | |
| Black Market | No | - | - | - | |
| BlackBank Market | Yes | 06.02.2014 | 17.05.2015 | Yes | |
| Blue sky | yes | 06.01.2014 | 28.09.2014 | Partly | Several months missing between scrapes |
| Cloud 9 | Yes | 11.02.2014 | 01.11.2014 | Yes | |
| Crypto Market | Yes | 19.02.2015 | 06.07.2015 | Yes | Strong variation on captured listings per scrape |
| Darknet Heroes League | Yes | 30.05.2015 | 04.07.2015 | Partly | |
| Diabolus/SR3 | Yes | 17.10.2014 | 05.07.2015 | Yes | |
| Dream Market | Yes | 09.01.2014 | 05.07.2015 | Partly | |
| East India Company | Yes | 28.04.2015 | 05.07.2015 | Yes | |
| Evolution | Yes | 21.01.2014 | 17.03.2015 | Yes | |
| Hansa | No | - | - | - | |
| House of Lions Market | No | - | - | - | |
| Hydra | Yes | 03.04.2014 | 27.10.2014 | Partly | Counterfeits are not captured in categories |
| Middle Earth Marketplace | Yes | 23.06.2014 | 05.07.2015 | Yes | |
| Mr Nice Guy 2 | Yes | 21.02.2015 | 04.07.2015 | - | Cannot inspect data (corrupted) |
| Nucleus Marketplace | Yes | 24.10.2014 | 07.07.2015 | Partly | Most of the scrapes were blocked (no content) |
| Outlaw Market | Yes | 09.01.2014 | 05.07.2015 | Partly | Almost exclusively drugs; newer listings without categorization |
| Pandora | Yes | 25.12.2013 | 05.11.2014 | No | |
| Pirate market | Yes | 25.01.2014 | 21.09.2014 | Yes | Big gaps in the scraped data |
| RoadSilk | Yes | 26.12.2013 | 15.01.2014 | Yes | |
| Silk Road 2.0 | Yes | 13.12.2013 | 06.11.2014 | Partly | |
| Silk Road 3.0 | No (see Diabolus) | - | - | - | |
| Silk Road Reloaded | Yes | 18.01.2015 | 05.07.2015 | Yes | No counterfeits found within categories |
| The Marketplace | Yes | 03.01.2014 | 09.11.2014 | Yes | |
| TheRealDeal | Yes | 16.04.2015 | 05.07.2015 | Partly | |
| Tochka | Yes | 05.02.2015 | 04.07.2015 | Partly | |
| Tom | Yes | 05.05.2014 | 17.12.2014 | Partly | 2 months of data missing |
| Topix | No | - | - | - | |
| Tor Bazaar | Yes | 02.02.2014 | 06.11.2014 | Partly | Almost exclusively drugs; Counterfeits not captured in categories |
| Valhalla | No | - | - | - | |

Table 10. List of considered markets; Cell properties: Yellow: Not in the archive; Orange: timeframe too short or data gaps too big; Blue: Products are not or only partly categorized; Grey: Data gaps too big or no counterfeits were found within the categories; Green: market included in analyses

**Appendix B: Categories included for keyword searches:**

"Other", "Fraud", "Electronics", "CustomListings", "Miscellaneous", "Accessories", "Fraud Related", "Jewelry", "Weight loss", "Forgeries", "Jewelry", "Listings", "Tobacco", "Market", "Hidden", "Precious metals", "Jewels & Gold", "Other Listings", "Abraxas", "Electronics", "Watches", "Accessories", "Clothing", "Sunglasses", "Cigarettes", "Jewelry", "Collectables", "Tobacco", "Metals Stones".

**Appendix C: Synonyms used for keyword search:**

"copies", "copy", "counterfeit", "replica", "fake", "clone", "deceit", "deception", "bum", "dummy", "facsimile", "gyp", "hoax", "humbug", "imitation", "imposture", "phony", "pseudo", "put-on", "reproduction", "sham", "simulacrum", "bogus", "junque", "likeness", "miniature", "lookalike", "xerox", "forge", "ditto", "dupe", "mimeo", "reduplication", "replication", "repro", "stat", "forgery", "forged", "spurious", "mock", "false", "unreal", "ungenuine", "falsified".

**Appendix D: Synonyms of authentic:**

"genuine", "authentic", "real", "valid", "original", "actual", "official".

**Appendix E: Keywords used to exclude listings:**

"tutorial", "template", "liscence", "ID", "refund", "scan", "scans", "COUPONS", "Licence", "License".

**Appendix F: Full lists of percentage counterfeits by OECD/EUIPO and IPO**

(OECD/EUIPO, 2019)

| Category | Share of custom seizures (%) | Share of seized value (%) |
|---|---|---|
| Footwear | 22.6 | 10.45 |
| Clothing, knitted or crocheted | 17.00 | 8.20 |
| Articles of leather | 13.55 | 11.60 |
| Electrical machinery and equipment | 12.25 | 10.75 |
| Watches | 5.70 | 22.75 |
| Optical, photographic and medical instruments | 5.15 | 4.10 |
| Perfumery and cosmetics | 3.50 | 4.95 |
| Toys | 2.75 | 4.65 |
| Jewellery | 1.85 | 5.85 |
| Machinery and mechanical appliances | 1.55 | 0.95 |
| Pharmaceutical products | 1.55 | 1.50 |
| Headgear | 1.45 | 0.30 |
| Other made up textile articles | 1.15 | 1.10 |
| Vehicles | 1.10 | 1.50 |
| Clothing, not knitted or crocheted | 1.00 | 0.95 |
| Plastics and articles thereof | 0.65 | 0.60 |
| Furniture, bedding, cushions, lamps etc. | 0.49 | 0.75 |
| Miscellaneous manufactured articles | 0.40 | 0.70 |
| Foodstuff | 0.40 | 0.56 |

Table 11. Full lists of counterfeit type shares by OECD/EUIPO (2019).

(IP Crime Group, 2015)

| Category | % |
|---|---|
| Optical Media | 39.85 |
| Tobacco | 28.15 |
| Clothing | 7.94 |
| Alcohol | 4.42 |
| Footwear | 2.77 |
| Circumvention | 2.01 |
| Cosmetics | 1.10 |
| Handbags | 1.44 |
| Software | 0.68 |
| E-Games | 0.66 |
| File-sharing | 0.78 |
| Other | 7.41 |
| Watches | 1.19 |
| Headphones | 0.44 |
| Electrical | 0.36 |
| Jewellery | 0.51 |
| Pharmaceuticals | 0.30 |

Table 12. Full lists of counterfeit type shares by IP Crime Group (2015).

**Appendix G: Product price differences for 10 products in each category between Darknet markets and the surface web.**

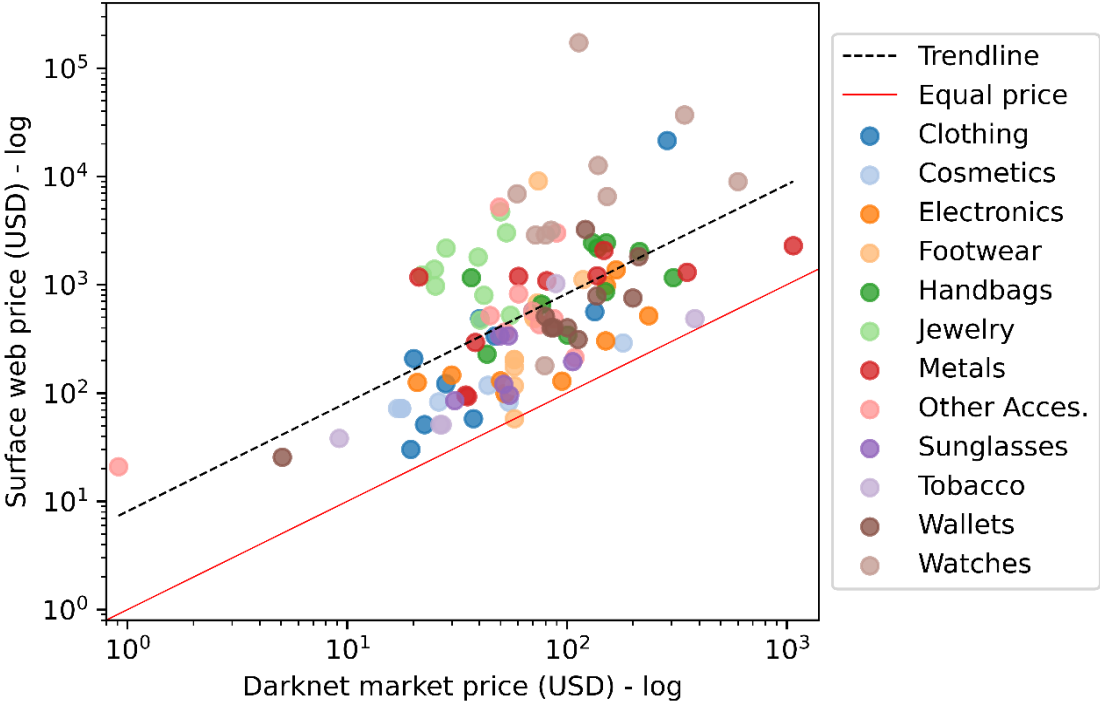

Figure 7. Product price differences for 10 products in each category between Darknet markets and the surface web.